# A THz Slot Antenna Optimization Using Analytical Techniques

*Shay ROZENBERG, Asher YAHALOM*

Department of Electrical & Electronic Engineering, Ariel University, Kiryat Hamada POB 3, Ariel, Israel

Shayrozenberg82@gmail.com, asya@ariel.ac.il

**Abstract.** *Slot antennas are very popular microwave antennas and slotted waveguides are used for high frequency radar systems. A thin slot in an infinite ground plane is the complement to a dipole in free space. This was described by H.G. Booker [2] who extended Babinet's principle from optics to show that the slot will have the same radiation pattern as a dipole such that the E and H fields are swapped. As a result, the polarization is rotated, so that radiation from vertical slot is polarized horizontally. In this work we show how analytical techniques can be used for optimization of THz slot antennas, the analysis is then corroborated by using a numerical simulation which validates the performance parameters predicted by the analytical technique.*

## 1. Introduction

Slot antennas are very popular microwave antennas and slotted waveguides are used for high frequency radar systems [1,2,3,4,5,6,7,8,9,10,11,12,13]. A thin slot in an infinite ground plane is the complement to a dipole in free space. This was described by H.G. Booker [2] who extended Babinet's principle from optics to show that the slot will have the same radiation pattern as a dipole such that the E and H fields are swapped. As a result, the polarization is rotated, so that radiation from vertical slot is polarized horizontally. For instance, a vertical slot has the same pattern as a horizontal dipole with the same dimensions and we are able to calculate the radiation pattern of a dipole. Thus, a longitudinal slot in the broad wall of a waveguide radiates just like a dipole perpendicular to the slot. By using this principle, it is easier to analyze slots antennas using the theory of dipole antennas. The slots are typically thin and 0.5 wavelengths long. The position of the slots affects the intensity of the transmitted power by being a direct cause to the impedance matching of the antenna, good impedance matching will cause maximum efficiency of the power transmission.

In this paper an antenna based on a slotted waveguide is analyzed analytically and designed for THz band (at frequencies of about 330 GHz).

In section 2, the antenna design is presented and the waveguide dimensions and the operation frequency are briefly discussed. First we describe a single slot on a rectangular waveguide and it is shown that it can be described using a model of a transmission line. By using this model, equations that relate the slot position to the transmitted power were developed. This model was expanded for an array of slots on a rectangular waveguide in order to design the slots antenna. In section 3, antenna based on slotted rectangular waveguide in the THz band was designed and simulated. The conclusions are given in section 4.

## 2. Antenna Design and Operation

In order to understand the slotted waveguide antenna, we will need to understand the fields within the waveguides first. In a waveguide, we are looking for solutions of Maxell's equations that are propagating along the guiding direction (the z direction). Thus, the electric and magnetic fields are assumed to have the form:

$$\tilde{\vec{E}}(x,y,z,t) = \tilde{\vec{E}}(x,y)e^{i\omega t - \beta z}$$
$$\tilde{\vec{H}}(x,y,z,t) = \tilde{\vec{H}}(x,y)e^{i\omega t - \beta z} \quad . \quad (1)$$

Where $\beta$ is the propagation wavenumber along the guide direction. The corresponding guide wavelength, is denoted by $\lambda_g = 2\pi/\beta$.

It is assumed that the reader is familiar with waveguide mode theory [1], and thus it will be stated without proof that the $TE$ field components in rectangular guide have the form:

$$\tilde{\tilde{H}}_z = \tilde{\tilde{H}}_{az} e^{-i\beta z}$$
$$\tilde{\tilde{E}}_t = \tilde{\tilde{E}}_{at} e^{-i\beta z} \quad . \quad (2)$$
$$\tilde{\tilde{H}}_t = \tilde{\tilde{H}}_{at} e^{-i\beta z}$$

Where $a$ is shorthand for the double index $mn$ and:

$$\tilde{\tilde{H}}_{az} = \cos\left(\frac{m\pi x}{a}\right) \cdot \cos\left(\frac{m\pi y}{b}\right) \hat{z}$$

$$\tilde{\tilde{E}}_{at} = \frac{-i\omega\mu}{k_c^2}\left(\hat{x}\frac{\partial \tilde{\tilde{H}}_{az}}{\partial y} - \hat{y}\frac{\partial \tilde{\tilde{H}}_{az}}{\partial x}\right) \quad . \quad (3)$$

$$\tilde{\tilde{H}}_{at} = \frac{-i\beta}{k_c^2}\left(\hat{x}\frac{\partial \tilde{\tilde{H}}_{az}}{\partial x} + \hat{y}\frac{\partial \tilde{\tilde{H}}_{az}}{\partial y}\right)$$

$$\beta = \sqrt{k^2 - k_c^2} = \sqrt{k^2 - \left(\frac{m\pi}{a}\right)^2 - \left(\frac{n\pi}{b}\right)^2}$$

The cutoff frequency of $TE_{mn}$ mode is expressed by the form:

$$f_{mn} = \frac{1}{2\pi\sqrt{\mu\varepsilon}}\sqrt{\left(\frac{m\pi}{a}\right)^2 + \left(\frac{n\pi}{b}\right)^2} \quad . \quad (4)$$

And the dominant mode fields, $TE_{10}$, are given by:

$$\tilde{\tilde{E}}_{10,y} = \frac{-i\omega\mu}{k_c^2}\frac{\pi}{a}\sin\frac{\pi x}{a}e^{-i\beta_{10}z}$$
$$\tilde{\tilde{H}}_{10,x} = \frac{i\beta}{k_c^2}\frac{\pi}{a}\sin\frac{\pi x}{a}e^{-i\beta_{10}z} \quad . \quad (5)$$
$$\tilde{\tilde{H}}_{10,z} = \cos\frac{\pi x}{a}e^{-i\beta_{10}z}$$

Figure 1 shows the waveguide description, the waveguide walls are perfectly conducting and the dimensions are chosen so that all modes except $TE_{10}$ are cut off.

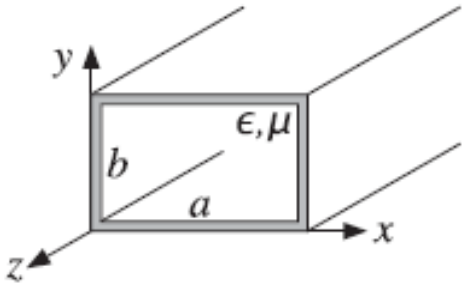

**Fig. 1.**  The waveguide description.

This information needs to be applied to scattering off a slot cut in one of the walls of the waveguide. If the waveguide is assumed to be infinitely long and a $TE_{10}$ mode is launched from $z = -\infty$, traveling in the positive z-direction, the incidence of this mode on the slot will cause backwards and forwards scattering of this mode and radiation into outer space is possible.

Figure 2 shows the geometry of the slot antenna. It is assumed that the waveguide walls have negligible thickness and are composed of perfect conductor. The slot is rectangular with length $2l$ and width $w$ where $2l \gg w$.

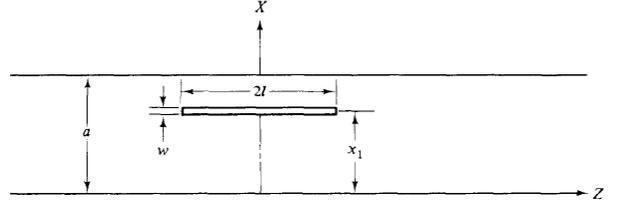

**Fig. 2.** An offset longitudinal slot in the upper broad wall of a rectangular waveguide.

The forward and backward scattering off this slot in the $TE_{10}$ mode will be [1]:

$$B_{10} = \frac{-\int_{x1-w/2}^{x1+w/2}\cos(\frac{\pi x}{a})dx\int_{-l}^{l}E_{1x}(z)e^{-i\beta_{10}z}dz}{\omega\mu\beta_{10}ab/(\pi/a)^2}$$

$$C_{10} = \frac{-(\pi/a)^2\cos(\frac{\pi x_1}{a})\int_{-l}^{l}V(z)e^{i\beta_{10}z}dz}{\omega\mu\beta_{10}ab} \quad . \quad (6)$$

respectively. In which $V(z) = w \cdot E_{1x}(z)$ is the voltage distribution in the slot. A slot in a large ground plane can be described as a two wire transmission line that fed at the central points, these "wires" are shorted at $z = \pm l$ as shown in figure 3.

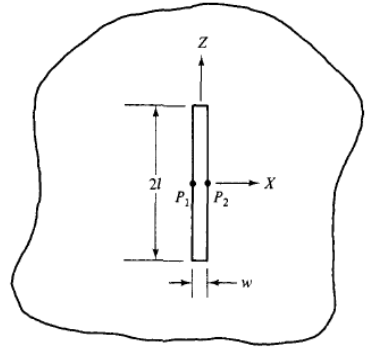

**Fig. 3.** Center-fed slot in a large ground plane.

Detailed analysis shows that the voltage distribution in the slot will be a symmetrical standing wave of the form:

$$V(z) = V_m \sin[k(l-|z|)] \quad . \quad (7)$$

When inserting (7) in (6) we will obtain:

$$B_{10} = C_{10} = \frac{2V_m}{\omega\mu(\beta_{10}/k)ab}(\cos\beta_{10}l - \cos kl)\cos(\frac{\pi x_1}{a}). \quad (8)$$

It is important to observe that the scattering off the slot being symmetrical, that is $B_{10} = C_{10}$. this implies that the slot is equivalent to shunt obstacle on a two wire transmission line. To see this, consider the situation suggested by figure 4. A transmission line of characteristic admittance $G_0$ is shunted at $z = 0$ by a lumped admittance $Y$.

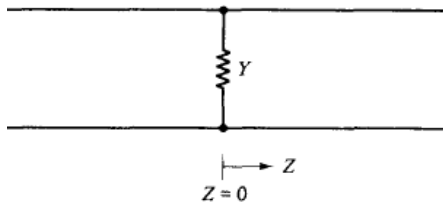

**Fig. 4.** A shunt obstacle on a two wire transmission line.

The voltage and current on the line are given by:

$$V(z) = Ae^{-i\beta z} + Be^{i\beta z}$$
$$z < 0$$
$$I(z) = AG_0 e^{-i\beta z} - BG_0 e^{i\beta z}$$
$$\quad (9)$$
$$V(z) = (A+C)e^{-i\beta z}$$
$$z > 0$$
$$I(z) = (A+C)G_0 e^{-i\beta z}$$

The boundary conditions are:

$$V(0^-) = V(0) = V(0^+)$$
$$I(0^-) = V(0)Y + I(0^+) \quad . \quad (10)$$

Which when inserted in (9), give:

$$B = C$$
$$\frac{Y}{G_0} = -\frac{2B}{A+B} \quad . \quad (11)$$

The usefulness of (11) lies in the fact that, by analogy, if one can find the ratio $-2B_{10}/(A_{10} + B_{10})$ for the slot, one can then say that the slot has equivalent normalized shunt admittance equal to that ratio.

The slot is said to be resonant if $Y/G_0$ is pure real. If we take $A_{10}$ to be pure real, it follows that the resonant conductance of the slot is given by:

$$\frac{G}{G_0} = -\frac{2B_{10}}{A_{10} + B_{10}}. \quad (12)$$

Where $B_{10}$ is perforce pure real also. What this implies is that, for a given displacement $x_1$ of the slot, it is assumed in (12) that the length $2l$ of the slot has been adjusted so that $B_{10}$ is either in phase with, or out of phase with, $A_{10}$.

The incident power is given by:

$$p_{inc} = \frac{1}{2}\text{Re}\int_{S1}(A_{10}\tilde{\vec{E}}_{10,t} \times A_{10}^*\tilde{\vec{H}}_{10,t}^*)\cdot\hat{z}dS_1$$
$$= \frac{\omega\mu\beta_{10}ab}{4(\pi/a)^2}A_{10}A_{10}^* \quad . \quad (13)$$

In like manner, one finds that the reflected and transmitted powers are:

$$p_{ref} = \frac{\omega\mu\beta_{10}ab}{4(\pi/a)^2}B_{10}B_{10}^*$$
$$p_{tr} = \frac{\omega\mu\beta_{10}ab}{4(\pi/a)^2}(A_{10}+C_{10})(A_{10}+C_{10})^* \quad . \quad (14)$$

If use is made of the information that $B_{10} = C_{10}$ and that all three amplitudes are pure real, then we can find expression to the power radiation:

$$power\ radiation = P_{inc} - P_{ref} - P_{tr}$$
$$power\ radiation = -\frac{\omega\mu\beta_{10}ab}{2(\pi/a)^2}B_{10}(A_{10}+B_{10}) \quad . \quad (15)$$

In [2] it is shown that the impedance relationship between slot and complementary dipole is:

$$R_{rad}^{dipole} \cdot R_{rad}^{slot} = \frac{\eta^2}{4}. \quad (16)$$

Where $\eta = 377\Omega$ and $R_{rad}^{dipole} = 73\Omega$, by using (16) we can find that $R_{rad}^{slot} = 486\Omega$. If we express the radiation resistance of the slot by $R_{rad}^{slot} = 486\Omega = \frac{\pi\eta/4}{0.609}$, we can find that:

$$P_{rad} = \frac{1}{2}\frac{V_m^2}{R_{rad}} = 0.609\frac{V_m^2 \cdot 2}{\pi\eta} \quad . \quad (17)$$

Thus:

$$-\frac{\omega\mu\beta_{10}ab}{2(\pi/a)^2}B_{10}(A_{10}+B_{10}) = 0.609\frac{V_m^2}{\pi\eta}. \quad (18)$$

If eliminated $V_m$ from (18) and (8), the result is:

$$\frac{G}{G_0} = [2.09\frac{(a/b)}{(\beta_{10}/k)}(\cos\beta_{10}l - \cos kl)^2 \cos^2(\frac{\pi x_1}{a})]. \quad (19)$$

When the substitution $x = x_1 - (a/2)$ is made and $kl = \pi/2$ is used, one obtains:

$$\frac{G}{G_0} = [2.09\frac{(a/b)}{(\beta_{10}/k)}\cos^2\left(\frac{\beta_{10}}{k}\frac{\pi}{2}\right)\sin^2\left(\frac{\pi x}{a}\right)]. \quad (20)$$

In which $x$ is the offset from the center line of the broad wall. Equation (20) indicates that the normalized conductance of a resonant longitudinal slot in the broad wall of a rectangular waveguide is approximately equal to a constant times the square of the sine of an angle proportional to its offset. From the point of view of the waveguide, the slot is shunt impedance across the transmission line, or equivalent admittance loading the transmission line. When the admittance of the slot (or combined admittance of all slots) equals the admittance of the guide, then we have matched transmission line and maximum power radiated.

In like manner, the circuit model of slotted waveguide antenna depicted in figure 5.

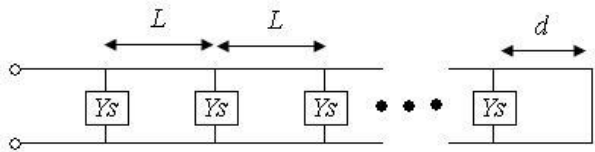

**Fig. 5.** Circuit model of slotted waveguide antenna.

The last slot is a distance $d$ from the end (which is shorted-circuited, as seen in figure 5), and the slot elements are spaced a distance $L$ from each other. The distance between the last slot and the end: $d$, is chosen to be a quarter-wavelength. Transmission line theory [1] states that impedance of a short circuit a quarter-wavelength down a transmission line is an open circuit, hence, figure 5 then reduces to:

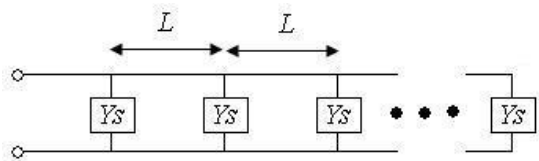

**Fig. 6.** Circuit model of slotted waveguide using quarter-wavelength transformation

The input admittance for an N element slotted array is:

$$Y = NY_S. \quad (21)$$

Sometimes the closed end is spaced $\frac{3}{4}\lambda_g$ for mechanical reasons; the additional half-wavelength is transparent. Spacing the slots at $\frac{1}{2}\lambda_g$ intervals in the waveguide is an electrical spacing of $180°$ - each slot is exactly out of phase with its neighbors, so their radiation will cancel each other. However, slots on opposite sides of the centerline of the guide will be out of phase, so we can alternate the slot displacement around the centerline and have a total phase difference of $360°$ between slots, putting them back in phase.

A photograph of a complete waveguide slot antenna is shown in figure 7. This example has 6 slots on each side for a total of 12 slots. The slots have identical length and spacing along the waveguide. Note how the slot position alternates about the centerline of the guide. The far wall of the waveguide has an identical slot pattern, so you can see through the slots.

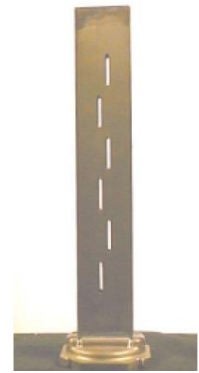

**Fig. 7.** Example of waveguide slot antenna.

A simple way to estimate the gain of slot antenna is to remember that is an array of dipoles. Each time we double the number of dipoles, we double the gain, or add 3 dB. The approximate gain formula is thus $Gain = 10\log(N)$ $dBi$ for N total slots. Ref [3] gives better formulas for the gain and beam width:

$$GAIN = 10\log\left(\frac{N\cdot\lambda_g/2}{\lambda}\right) dBi$$

$$Beamwidth = 50.7\left(\frac{\lambda}{N/2 \cdot \lambda_g/2}\right) [°] \quad (22)$$

It is known from [4] that the slot width should be taken as $w = \lambda_g / 20$. But, measurements taken by Stegen in [5] were based on a slot width of 1.5875 mm in WR-90 waveguide. For other waveguide sizes, the slot width should be scaled accordingly.

# 3. Slot Antenna Design and Simulation

We can summarize the design procedure for waveguide slot antenna as follows:

1. Choose the number of slots required for the desired gain and beam width.

2. Choose a waveguide size appropriated for the operating frequency.

3. Calculate the wavelength in the waveguide at the operating frequency.

4. Determine the slot dimensions, length and width appropriated for the operating frequency.

5. Determine the slot position from centerline for normalized admittance of 1/N, where N is the number of slots in both walls of the waveguide.

6. Space $\tfrac{1}{4}\lambda_g$ or $\tfrac{3}{4}\lambda_g$ between the center of the last slot and the end of the waveguide.

For a THz slot antenna design at operation frequency of 330 GHz. The waveguide dimensions are chosen so that all modes except $TE_{10}$ are cut off. The waveguide is WR3 with the dimensions of $0.864 \times 0.432 \,[mm^2]$.

The wavelengths are:

$$\lambda = \frac{c}{f} = 0.909 [mm]$$
$$\lambda_c = \frac{2\pi}{k_c} = \frac{2\pi}{\frac{m\pi}{a}+\frac{n\pi}{b}} = 2a = 1.728 [mm] \quad (23)$$
$$\lambda_g = \frac{1}{\sqrt{\frac{1}{\lambda^2}-\frac{1}{\lambda_c^2}}} = 1.07 [mm]$$

By choosing 256 slots we will obtain:

$$GAIN = 10\log\left(\frac{N \cdot \lambda_g/2}{\lambda}\right) = 10\log\left(\frac{256 \cdot 1.07/2}{0.909}\right) = 21.8\, dBi \quad (24)$$

$$Beamwidth = 50.7\left(\frac{\lambda}{N/2 \cdot \lambda_g/2}\right) = 50.7\left(\frac{0.909}{256/2 \cdot 1.07/2}\right) = 0.7[°]$$

So the resolution at 10 m is 11.74 [cm].

The slot dimensions are:

$$slot\_length = 2l = \frac{\lambda}{2} = 0.454 [mm]$$
$$slot\_width = w = \frac{\lambda_g}{2} = 0.0535 [mm] \quad (25)$$

The offset from the center line of the broad wall is:

$$\frac{G}{G_0} = [2.09\frac{(a/b)}{(\beta_{10}/k)}\cos^2\left(\frac{\beta_{10}}{k}\frac{\pi}{2}\right)\sin^2(\frac{\pi x}{a})]$$
$$\frac{1}{N} = [2.09\frac{(a/b)}{(\beta_{10}/k)}\cos^2\left(\frac{\beta_{10}}{k}\frac{\pi}{2}\right)\sin^2(\frac{\pi x}{a})]$$
$$N \cdot [2.09\frac{(a/b)}{(\beta_{10}/k)}\cos^2\left(\frac{\beta_{10}}{k}\frac{\pi}{2}\right)\sin^2(\frac{\pi x}{a})] = 1$$
$$\sin(\frac{\pi x}{a}) = \frac{1}{\sqrt{N[2.09\frac{(a/b)}{(\beta_{10}/k)}\cos^2\left(\frac{\beta_{10}}{k}\frac{\pi}{2}\right)]}}$$
$$x = \frac{a}{\pi}\arcsin\frac{1}{\sqrt{N[2.09\frac{(a/b)}{(\beta_{10}/k)}\cos^2\left(\frac{\beta_{10}}{k}\frac{\pi}{2}\right)]}}$$
$$x = 0.033 [mm] \quad (26)$$

Figures 8 and 9 shows a model performed using CST Microwave Studio according to the analytical design, the simulation result for the radiation pattern, gain and beam width depicted in figures 10-14.

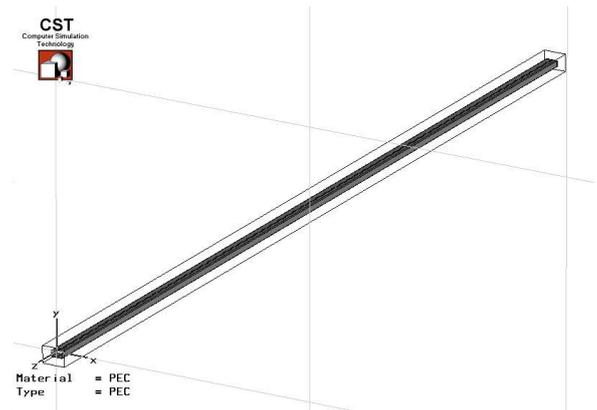

**Fig. 8.** Slot antenna model using CST Microwave Studio.

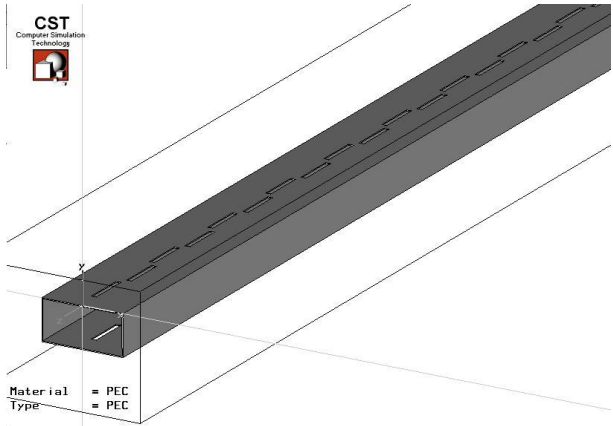

**Fig. 9.** Slot antenna model using CST Microwave Studio (zoomed in).

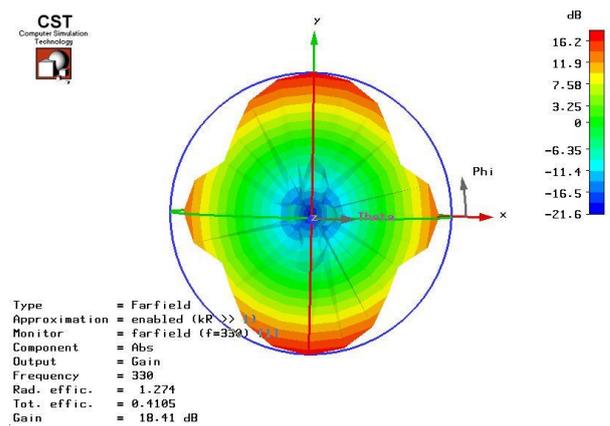

**Fig. 12.** Radiation pattern simulation result (z-axis view).

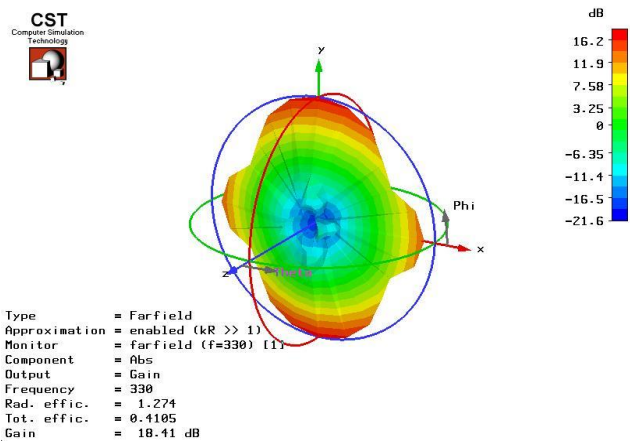

**Fig. 10.** Radiation pattern simulation result.

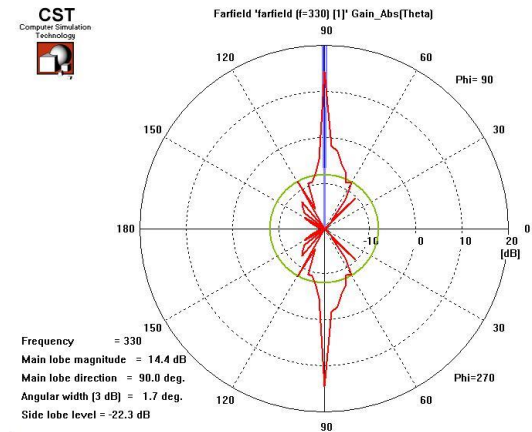

**Fig. 13.** Azimuth radiation pattern simulation result.

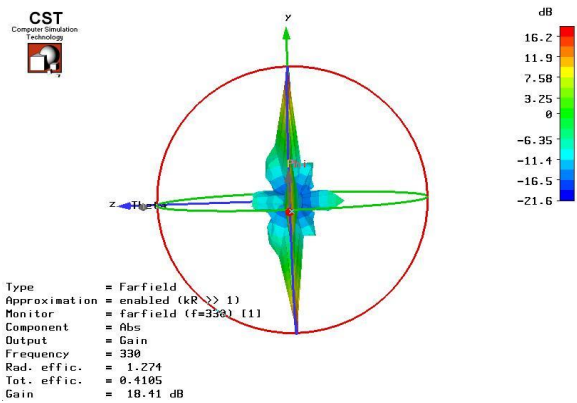

**Fig. 11.** Radiation pattern simulation result (x-axis view).

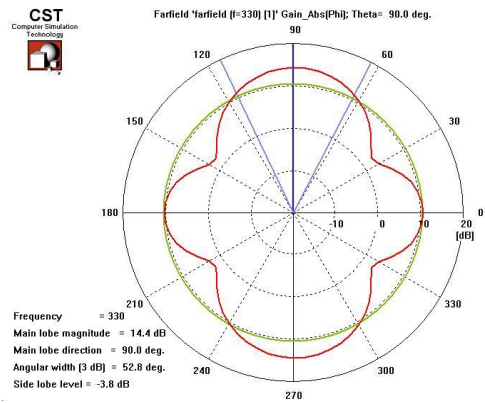

**Fig. 14.** Elevation radiation pattern simulation result.

The simulation results show that the antenna gain is 18.4 dBi and the azimuth beam width is $1.7°$ compared to 21.8 dBi and $0.7°$ in the analytical design.

There is a small difference between the analytical and simulation results, the difference is due to the simplifying assumptions that were taken in the analytical design (wall thickness was ignored; waveguide length was taken as infinity etc.). Despite the difference between analytical and the simulation results, good approximation of antenna's performance, antenna's pattern, beamwidth and gain can be made.

# 4. Conclusion

In this paper, slot antenna based on rectangular waveguide has been presented and investigated. It has been shown that slot antenna in the high frequency regime can be analyzed and optimized by an analytical model this was verified using a simulation software. To optimize the design and obtain the required antenna's pattern and performance it is first recommended to use an analytical model of the antenna. Using the analytical design information one can perform a simulation. Then, according to the simulation results, modify the parameters until the optimum results are obtained.

# About the Authors

**Shay ROZENBERG** is a Radar System Engineer. He studied towards his BSc degree in electrical engineering during 2004 – 2006 in Ariel University. During the years 2009 – 2014 he studied toward an MSc in RF Engineering in Ariel University under the supervision of Professor Asher Yahalom.

**Asher YAHALOM** is a Full Professor in the Faculty of Engineering at Ariel University and the Academic director of the free electron laser user center which is located within the University Center campus. He was born in Israel on November 15, 1968, received the B.Sc., M.Sc. and Ph.D. degrees in mathematics and physics from the Hebrew University in Jerusalem, Israel in 1990, 1991 and 1996 respectively. From 1994 to 1998 Asher Yahalom worked with Direx Medical System on the development of a novel MRI machine as a head of the magneto-static team. Afterwards he consulted the company in various mathematical and algorithmic issues related to the development of the "gamma knife" - a radiation based head surgery system. In the years 1998-1999 Asher Yahalom joined the Israeli Free Electron Laser Group both as postdoctoral fellow and as a project manager, he is a member of the group ever since. In 1999 he joined the College of Judea & Samaria which became at 2007 Ariel University Center. During 2005-2006 on his first sabbatical he was a senior academic visitor at the institute of astronomy in Cambridge. During his second sabbatical in the years 2012-2013 he was a visiting fellow of the Isaac Newton Institute for Mathematical Sciences also in Cambridge UK. Since 2013 Asher Yahalom is a full professor and since 2014 he is the head of the department of electronic & electrical engineering. Asher Yahalom works in a wide range of scientific & technological subjects ranging from the foundations of quantum mechanics to molecular dynamic, fluid dynamics, magnetohydrodynamics, electromagnetism and communications.